\documentclass[fleqn,usenatbib]{mnras}

\usepackage[T1]{fontenc}

\DeclareRobustCommand{\VAN}[3]{#2}
\let\VANthebibliography\thebibliography
\def\thebibliography{\DeclareRobustCommand{\VAN}[3]{##3}\VANthebibliography}

\usepackage{graphicx}	
\usepackage{amsmath}	
\usepackage{amssymb}	
\usepackage{multirow}
\usepackage{todonotes}
\usepackage{multicol}
\usepackage{marginnote}

\usepackage{lineno}

\usepackage[normalem]{ulem}

\title[LFQPOs in accretion onto BH]{Low Frequency Quasi Periodic Oscillations and Shocks in Accretion onto Black Hole}

\author[Singh, Mondal \& Garofalo]{
Chandra B. Singh,$^{1,4}$\thanks{Email: chandrasingh@ynu.edu.cn}
Santanu Mondal,$^{2,4}$\thanks{E-mail: santanuicsp@gmail.com}
David Garofalo$^{3}$
\\
$^{1}$ South-Western Institute for Astronomy Research, Yunnan University, University Town, Chenggong, Kunming 650500,    \\People's Republic of China\\
$^{2}$Indian Institute of Astrophysics, II Block, Koramangala, Bengaluru 560 034, India\\
$^{3}$Department of Physics, Kennesaw State University, Marietta, GA 30060, USA\\
$^{4}$Equal first authors
}

\pubyear{2020}

\begin{document}
\label{firstpage}
\pagerange{\pageref{firstpage}--\pageref{lastpage}}
\maketitle

\begin{abstract}
Low-frequency quasi-periodic oscillations (LFQPOs) have been routinely observed in black hole X-ray binaries (BHXRBs). These LFQPOs can be explained by axisymmetric shock oscillation in accretion flow around a rotating black hole. We address the physical origin of Type-C LFQPOs in BHXRBs observed by the Rossi X-ray Timing Explorer satellite considering a minimum number of free parameters, namely, specific energy and specific angular momentum of the infalling matter for a given set of BH mass and spin parameter. We apply the solution for a large number of BH candidates to further strengthen the scenario of an anti-correlation between the QPO frequency and the location of the shock. Our study also confirms that Compton cooling can be sufficient to explain the observed QPOs.
\end{abstract} 

\begin{keywords}
{black hole physics --- accretion --- accretion discs --- hydrodynamics --- shock waves --- binaries: close}
\end{keywords}

\section{Introduction}
Black hole X-ray binaries (BHXRBs) consist of a stellar-mass black hole (BH) fed by material supplied from a companion star. The accretion flow close to the central BH is quite hot and the radiation is likely to emit in the X-ray regime. Usually, BHXRBs are likely to be transient and undergo changes in ``states'' like quiescence and outbursts. There are mainly four types of spectral states from hard state (HS) to soft state (SS) through intermediate states (see Remillard \& McClintock 2006). Low hard state(LHS) is when hot Comptonized emission dominates over the cold thermal emission while in the high soft state(HSS) thermal emission is the dominant one. For intermediate states, namely hard intermediate (HIMS) and soft intermediate (SIMS), both non-thermal, as well as thermal components are significant. The fourth type of state is the anomalous state or ultra-luminous state (ULS) which is similar to SIMS but at higher luminosity and variability (Belloni 2010 and references therein). \\

The X-ray emissions from BHXRBs are likely to show temporal features like quasi-periodic oscillations (QPOs), varying from a few mHz to few tens of kHz range. QPOs usually become noticeable when the Fourier transform of the light curve is taken and the power density spectrum is produced. QPOs in BHXRBs are classified into two domains: low-frequency QPOs (LFQPOs) and high-frequency QPOs (HFQPOs) with centroid frequency $\le$ 30 Hz and $\ge$ 60 Hz respectively (Belloni 2010 and references therein). Further, there are three types of LFQPOs in BHXRBs, namely, type- A, B, and C (Casella et al. 2005). Type-A is the least common while type-C is the most common among observed LFQPOs. Although type-C LFQPOs are present in all possible spectral states, usually seen by end of LHS and HIMS with frequency $\nu$ in the range of few mHz and 10 Hz (Motta et al. 2012).  

The physical origin of LFQPOs in general and type-C, in particular, have been an active area of research. Some works propose different kinds of instabilities due to magneto-acoustic waves (Titarchuk et al. 1998; Cabanac et al. 2010), spiral density waves (Varniere \& Tagger 2002; Varniere et al. 2012), etc. While others consider general relativistic effects like Lense-Thirring precession (Stella \& Vietri 1999; Ingram et al. 2009). Even though several models are available which can explain the observed QPOs, they do not explain the observed states in a way that connects with the QPOs. A comprehensive explanation must physically connect the timing and spectral properties.
Moreover, Marcel \& Neilsen (2021) recently pointed out that the most popular precessing inner flow model (Ingram et al. 2009) does not take into account realistic accretion flow properties, rather considers a specific structure of solid-body for precession, which is unlikely to be responsible for producing Type-C LFQPOs around BHXRBs. Therefore, it is suggestive and worth considering models which can explain both properties together.

Accretion onto the BH is necessarily transonic and multiple sonic points can exist. In such a scenario, standing shocks are likely to form depending on the choice of the flow parameters. The post-shock region formed by low angular momentum sub-Keplerian flow is likely to have all required properties of a corona around BH (Chakrabarti 1989; Chakrabarti \& Titarchuk 1995, hereafter CT95). The accretion flow with two components: sub-Keplerian halo and Keplerian disc can successfully address the spectral as well as temporal properties of BH (CT95, Chakrabarti et al. 2015, Debnath et al. 2014). According to this model, the shock, which occurs at the coronal boundary layer, connects both spectral and temporal properties by changing its size, strength along with changes in mass accretion rates and by its oscillation due to the presence of cooling (Molteni et al. 1996, Garain et al. 2014). The physical origin of LFQPOs can be due to the resonance condition when the timescale of cooling due to Comptonisation matches with the timescale of infall in the coronal region. In this model, the post-shock corona region acts as base of the outflows (Chakrabarti 1999; Singh \& Chakrabarti 2011).

It is likely that BHs are spinning in nature (McClintock et al. 2014 for a review) and can determine the nature of standing shocks: whether form closer or farther from the BH and its strength. It is also possible that BH properties like its mass and spin can be inferred from QPO properties in observations (Zhang et al. 1997; Cui et al. 1998). In this work, we decipher the physical origin of LFQPOs, especially commonly observed type-C, and address all available QPO frequencies for a large number of BHXRBs samples observed by Rossi X-ray Timing Explorer (RXTE) (Motta et al. 2015). The mass and the spin parameters for the individual sources are taken from the literature. Our goal in this study is to apply the well-known shock oscillation model to fit and explain the origin of observed LFQPOs and further establish that Compton cooling can be sufficient to explain them. Such a study has not been performed before to the best of our knowledge.

The paper is organized in the following way: in the next section, we present the governing equations and methodology to evaluate QPO frequencies. In \S 3, we discuss results explaining observed Type-C LFQPOs (hereafter, QPOs) in BHXRBs. 
Finally, in \S 4, we briefly summarize our results and conclude.

\section{Governing equations and methodology}

Matter accreting onto a black hole forms a disc-like structure. In our present solution, we consider a thin, axisymmetric, low angular momentum and adiabatic flow around a rotating BH, where the flow is in vertical equilibrium. The fluid equations are time-independent for inviscid accretion flows. We have integrated the flow from the inner sonic point to the outer sonic point. In order to have a shock, the flow must be supersonic when passing through the inner sonic point, i.e. transonic (see Chakrabarti 1989, hereafter C89). The complete solution of the stationary model requires the dimensionless equations of the energy, the angular momentum, and the mass conservation supplied by the transonic conditions at the critical points and the Rankine-Hugoniot conditions at the shock. Presently, however, we have considered dissipation at the post-shock region. The sonic point conditions and shock location are derived following the similar procedures presented in C89.
The BH mass, flow velocity, and length scales are in units of $M_\odot$, the velocity of light c and $r_g=GM_{\rm BH}/c^2$ respectively for the rest of the paper.
Here, we put the model equations in absence of turbulence similar to those presented in Mondal (2020) for the sake of completeness. Similar sets of equations and analysis procedure can also be seen in Das et al. (2010). 
The dimensionless energy conservation equation of the flow in pseudo-Kerr (PK) geometry can be written as (C89),

\begin{equation}
\varepsilon=\frac{v^2}{2}+\frac{1}{\gamma-1}{a_s}^2+\Phi_{\rm PK},
\end{equation}
where $\gamma$, $v$ and $a_s$ are the adiabatic index, the radial velocity of the flow,
and the adiabatic sound speed defined by $\sqrt{\gamma P/\rho}$. Here $P$ is the isotropic pressure and $\rho$
is the gas density. We consider the value of $\gamma=4/3$, constant throughout the flow.
The PK potential ($\Phi_{\rm PK}$), introduced by Chakrabati \& Mondal (2006), 
is given by,
\begin{equation}
\Phi_{\rm PK}=-\frac{B+\sqrt{B^2-4AC}}{2A},
\end{equation}
where,

\begin{equation} 
A=\frac{\alpha^2 l^2}{2r^2sin^2\theta}, 
\end{equation}

\begin{equation}
B=\frac{\alpha^2 \omega l}{sin^2\theta}+\frac{2al}{r^3sin\theta}-1, 
\end{equation}
and
\begin{equation}
C=1-\frac{1}{r-r_0}+\frac{2a\omega}{rsin\theta}+\frac{\alpha^2 \omega^2 r^2}{2sin^2\theta}. 
\end{equation}

Here, r, $\theta$, $l$, and $a$ are the radial distance in
spherical coordinate, disc inclination angle, the specific angular momentum, and the spin parameter defined by specific spin angular momentum of a rotating BH. Here,
$r_0=0.04+0.97a+0.085a^2$, and $\alpha^2=(r^2+a^2-2r)/(r^2+a^2+2a^2/r)$,
is the redshift factor. The above potential can mimic the general relativistic considerations up to a spin parameter ($a$) $\leq0.8$.\\ 

As the matter comes closer to the BH, centrifugal force increases rapidly by $\sim l^2/r^3$, produces a centrifugal
barrier close to the BH, and balances with the gravitational attraction.
Accreting matter piles up behind this layer and ends up forming a shock. For our purposes, we will use the Rankine-Hugoniot shock  
conditions (Landau \& Lifshitz 1959) to find the location of the shock.
The conservation of energy at the shock in presence of dissipation (cooling) is given by,
\begin{equation}
\varepsilon_{+}=\varepsilon_{-} - \Delta \varepsilon,
\end{equation}
where $\Delta \varepsilon$ is the cooling due to thermal Comptonisation. 
\begin{table*}
\centering
\caption{Scaling parameters and fits statistics for \autoref{fig:CoolingVsQPO}  \label{table:FitStat}}
\begin{tabular}{lcccccr}
\hline
Candidates        &$f_{\rm 0}$ & TS &p-value &$\alpha$ & TS &p-value \\
\hline
MAXI~J1543-564    &$0.017\pm0.001$&24.82&1.4e$-$4&$0.49\pm0.05$&10.32&0.002  \\
MAXI~J1836-194    &$0.014\pm0.001$&23.29&1.7e$-$4&$0.72\pm0.06$&11.34&0.002 \\
H~1743-322        &$0.04\pm0.01$&5.17&0.001&$0.46\pm0.09$&5.2&0.001 \\
GRO~J1655-40      &$0.010\pm0.001$&6.0&1.9e$-$5&$0.18\pm0.029$&6.0&1.9e$-$5 \\
\hline
\end{tabular}
\end{table*}

For the estimation of $\Delta \varepsilon$, we follow the steps below: (i) we consider the cooling from Mondal et al. (2017) for four BH candidates namely, MAXI~J1543-564, MAXI~J1836-194, H1743-322, and GRO J1655-40. The details of the cooling estimation method can be seen in Mondal et al. (2015).
The $\Delta \varepsilon$ used has been estimated using the X-ray spectra observed by RXTE/PCA in the range 2.5-25 keV. It is the area under the curve in the luminosity-energy plot, therefore, uncertainty in its estimation is also on the same order as in luminosity estimation, (ii) cooling is introduced in our hydrodynamic code which solves stationary, conservation equations of inviscid fluid in 1.5 dimension, namely the height is taken into account for the low angular momentum radial flow and the sound speeds before ($a_-$) and after ($a_+$) the shock location are found, (iii) considering energy dissipation across the shock in a radiatively-inefficient flow, the energy loss mostly can occur through thermal Comptonization and is proportional to the temperature (and hence the sound speed) difference in the pre-shock and post-shock regions, $\Delta \varepsilon=f_c n (a_+^2-a_-^2)$ presented in Das et al. (2010); Singh \& Chakrabarti (2011), we estimate the $f_c$ values corresponding to each observed QPO frequency, here, n is the polytropic index of the flow, is written as $1/(\gamma-1)$, (iv) thereafter the $f_c$ profile is fitted as a function of $\nu_{\rm QPO}$ with a power-law function ($f_0 \nu_{\rm QPO}^{-\alpha}$). In \autoref{table:FitStat} we summarize the power-law fitted parameters and their error bars. We also tabulate the fit statistics, both t-statistics and p-value obtained from the fits for all four sources, (v) we find that the function $f_c=0.02\nu_{\rm QPO}^{-0.6}$ fits the observed $f_c$ profile for all four sources within 50\% upper and lower limit of the best fitted power-law curve (green solid line in \autoref{fig:CoolingVsQPO}). 

The current cooling is estimated from the thermal spectra of the sources, which might change if the non-thermal component and some other cooling mechanisms are taken into account. The four sources cover the full range of black hole masses in BHXRBs. Furthermore, the QPO frequencies observed for these four sources cover the range of the QPOs observed in all BHXRBs. The functional behavior proposed for fc is not necessarily universal, as it might require further study considering a larger population of black holes, including intermediate-mass and supermassive black holes. However, we anticipate that, as the current profile addresses both the mass of the low mass black hole range and QPO frequency range, it can fit those properties of all other BHXRBs.

It should be noted that the 50\% upper and lower bounds are chosen not from the statistical point of view, rather from observational findings. It has been verified in Chakrabarti et al. (2015) that once the resonance condition satisfies (cooling timescale $\sim$ infall timescale), it gets locked for a couple of  days or weeks and the QPOs can be observed when the resonance condition varies 50\% above and below 1. We apply that finding in our current study, (vi) for all sources presented in this paper we use the fitted function in (v), (vii) for all possible sets of $\varepsilon$ and $l$, flow generates shocks, which gives QPO frequency and the solution keeps searching a suitable location of the shock, which can generate the observed frequency. We repeat the above steps for all BHs considered here.
Here, ``-'' and ``+'' signs stand for pre- and post-  shock quantities.
The mass conservation equation at the shock is given by,
\begin{equation}
\dot{M_{+}}=\dot{M_{-}}.
\end{equation}
The pressure balance condition is given by,
\begin{equation}
W_{+} +\Sigma_{+} v_{+}^2 = W_{-} + \Sigma_{-} v_{-}^2.
\end{equation}

where W, $\Sigma$, and $v$ are the vertically integrated pressure, density and radial velocity of the flow. 
After solving a few steps, one can get the pre-and post-shock Mach number relation (C89) at the shock, as below:
\begin{equation}
C_{\rm shk}=\frac{[2+M_+^2(3\gamma-1)]^2}{M_+^2[2+(\gamma-1)M_+^2]}=\frac{[M_-^2(3\gamma-1)+2]^2}{M_-^2[2+(\gamma-1)M_{-}^2-\zeta]},
\end{equation}
where, $\zeta=\frac{2 \Delta \varepsilon (\gamma-1)}{a_{s-}^2}$. Due to piling up of the matter in post-shock flow, it becomes more turbulent, thus the shock invariant quantity indeed modifies if one takes into account the effect of turbulent pressure, which also modifies the QPO frequency significantly (Mondal 2020).
 
From Eq.~9, the pre-and post-shock Mach number can be written as,
\begin{equation}
        M_{\mp}^2=\frac{-Y\pm\sqrt{Y^2-4XZ}}{2X},
\end{equation}
where, $X=(3\gamma-1)^2-C_{\rm shk}(\gamma-1)$, $Y=4(3\gamma-1)-2C_{\rm shk}-\frac{2\Delta \varepsilon(\gamma-1) C_{\rm shk}}{a_{s-}^2}$, $Z=4$,
in terms of mach invariant quantity, $C_{\rm shk}$ at the shock. Here, $M_{\mp}$ is the Mach number of the pre-and post-shock flow.
We follow the same mathematical procedure and solution technique as discussed in C89 to find the shock location ($X_s$ in $r_g$ unit). It is to be noted that for studying accretion dynamics around black hole unlike particle dynamics (wherein one can choose a location in an ad-hoc manner and no information is available for the correct treatment of inner and outer boundary conditions), it is required that we consider at least two free parameters, $\varepsilon$ and $l$,  which not only determine $X_{s}$ but also the global flow solution containing multiple critical points and all necessary variables like density, pressure, and velocity along the flow.
As we consider that the QPOs can be originated due to the oscillation of the shock and the frequency is
found to be related to the infall time ($t_{\rm infall}$) $\sim$
$X_s/v_+$ $\sim$ $R X_s/v_-$ and $\nu_{\rm QPO}=1/t_{\rm infall} = \nu_0 v_-/R X_s$ (Molteni et al. 1996; Chakrabarti et al. 2005), where $\nu_0$ is $c/r_g$.
Later it was verified by numerical simulation (Garain et al. 2014) and also from observation (Chakrabarti et al. 2015).

\section{Results}

\begin{table*}
\centering
\caption{OBSERVED AND THEORETICAL PARAMETERS USED FOR ESTIMATING THEORETICAL LFQPOs.  \label{table:results}}
\begin{tabular}{lccccr}
\hline
Candidates        &$M_{\rm BH}$ &$a$ &$\varepsilon$ &$l$&Refs. \\
\hline
4U~1543-47        &9.4    &0.55   &1.0001-1.0025 &3.036-3.058&S06, M09  \\
GRO~J1655-40      &6.0    &0.5    &1.0001-1.0007 &3.110-3.338&S03, AK01 \\
MAXI~J1535-571    &8.9    &0.67   &1.0001-1.0002 &2.984-3.062&S19, X18 \\
Swift~J1753.5     &5.5    &0.75   &1.0001 &2.896-2.942&N14, D17, R09 \\
XTE~J1550-564     &9.1    &0.34   &1.0001-1.0017 &3.298-3.534&O11, M14 \\
XTE~J1650-500     &7.3    &0.8    &1.0001-1.0009 &2.814-2.832&SS08  \\
XTE~J1817-330     &6.0    &0.22   &1.0001 &3.42-3.426&S07, F17 \\
XTE~J1859+226     &6.5    &0.4    &1.0001-1.0015 &3.248-3.394&C11, F17  \\
\hline
H~1743-322        &11.2   &0.4    &1.0001 &3.27-3.38&M17, TK18 \\
XTE~J1752-233     &10     &0.52   &1.0001  &3.136-3.202&C20, R11 \\
\hline
\end{tabular}
\\
\noindent{Refs: S06-Shafee et al. 2006, M09-Miller et al. 2009, S03-Shahbaz et al. 2003, AK01- Abramowicz \& Kluzniak 2001, S19- Shang et al. 2019, X18- Xu et al. 2018, N14- Nuestroev et al. 2014, D17- Debnath et al. 2017, R09- Reis et al. 2009, O11- Orosz et al. 2011, M14- Motta et al. 2014, SS08- Slany \& Stuchlik 2008, S07- Sala et al. 2007, F17-Franchini et al. 2017, C11-Corral-Santana et al. 2011, M17- Molla et al. 2017, TK18- Tursunov \& Kolos 2018, C20- Chatterjee et al. 2020, R11- Reis et al. 2011.}
\end{table*}

\begin{figure*}
\centering{
\includegraphics[width=0.75\linewidth]{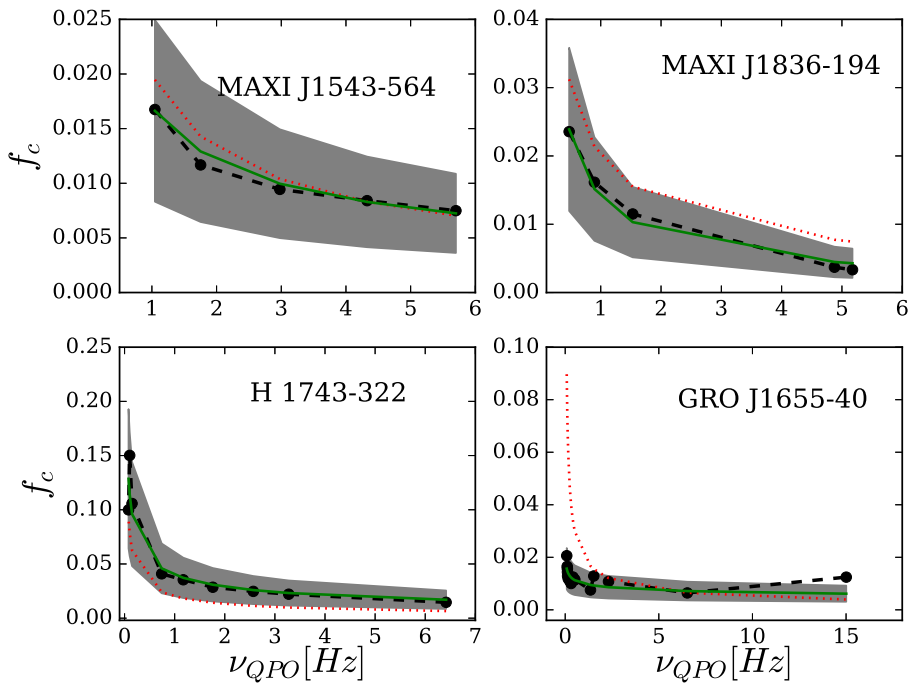}}
\caption{Functional dependence of cooling factor, $f_{c}$, on QPO frequency,$\nu_{\rm QPO}$, for BHXRBs. Black points represent observational data points, the green solid lines are fitted power-law function, the grey shaded region includes the 50\% range of theoretically fitted (green) values followed by the satisfaction of resonance condition. The red dotted curve shows the generalized function used in this paper to estimate QPO frequencies for other BHs.
\label{fig:CoolingVsQPO}}
\end{figure*}

Table~\ref{table:results} summarizes properties of BHXRBs, namely mass ($M_{\rm BH}$) and spin ($a$), which exhibit QPOs in RXTE observations and the best possible flow parameters (in columns 4 and 5) in our theoretical solution which can address the physical origin of QPOs. It can be noticed that even a narrow range of specific energy ($\varepsilon$) and specific angular momentum ($l$) can successfully explain QPOs in general. Since we do not have $\varepsilon$ and $l$ beforehand apart from QPO frequency from observation, we solve the flow equations for all possible values of those two basic quantities that can produce shock, thus the QPOs. Therefore it is likely to have degeneracy in the solution as we are exploring the whole parameter space. Furthermore, QPOs are observed for a couple of days to weeks, for some sources it can be months, therefore by that time flow can change its topology due to the presence of mass outflow from the disc surface, which can carry both energy and angular momentum from the flow (Blandford \& Begelman 1999). This implies that for a given source it is not expected to have the same $\varepsilon$ and $l$ throughout the QPO observation period. Therefore the degeneracy in the solution we are getting has physical implications both in accretion onto BHs and the origin of QPOs. However, this degeneracy is higher in the case of LFQPOs as the area of shock formation is larger and the QPO frequencies are less sensitive to the spin of the BH.
Fig.~\ref{fig:CoolingVsQPO} shows the variation of cooling factor $f_{c}$ with QPO frequency $\nu_{\rm QPO}$ for some typical BHXRBs: MAXI J1543-564, MAXI J1836-194, H1743-322, and GR0J1655-40. These BHXRBs cover the possible range of $M_{\rm BH}$ and $a$. For the given value of observed  $\nu_{\rm QPO}$, typical values of $f_{c}$ were estimated using the methodology described in the previous section. Based on Fig.~\ref{fig:CoolingVsQPO}, a functional dependence of $f_{c}$ on $\nu_{\rm QPO}$ can be written as $f_{c} \propto 0.02 \nu_{\rm QPO}^{-0.6}$. It can be seen from four different cases that the fitting relation can be safely applied for $\nu_{\rm QPO} \ge 0.5$~Hz. This indicates that the amount of cooling in the post-shock region determines the QPO properties. The higher the cooling, the lower is the QPO frequency. Such functional relation can be verified in the case of other BHXRBs as well and a general trend is established. As the cooling is estimated from integrating the Comptonisation part of the spectrum and the $f_c$ is estimated using the same cooling and pre-and post-shock sound speeds solving hydrodynamic equations, it can make a bridge between theory and observation in finding QPOs. Such attempt has not been done before in other theoretical studies.\\

\begin{figure*}
\centering{
\hspace{-1.2cm}
\includegraphics[width=0.37\linewidth]{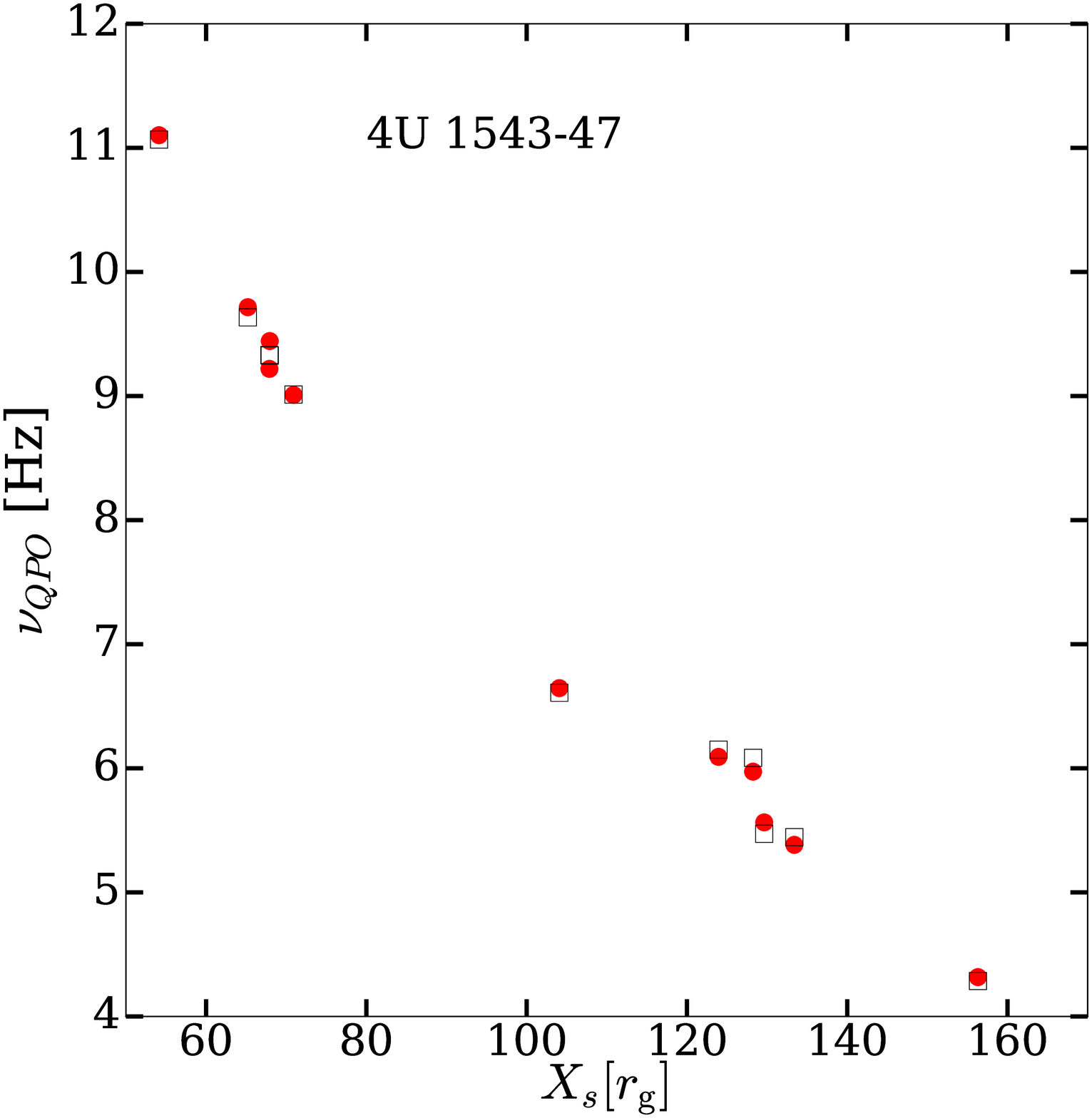}
\hspace{-0.6cm}
\includegraphics[width=0.37\linewidth]{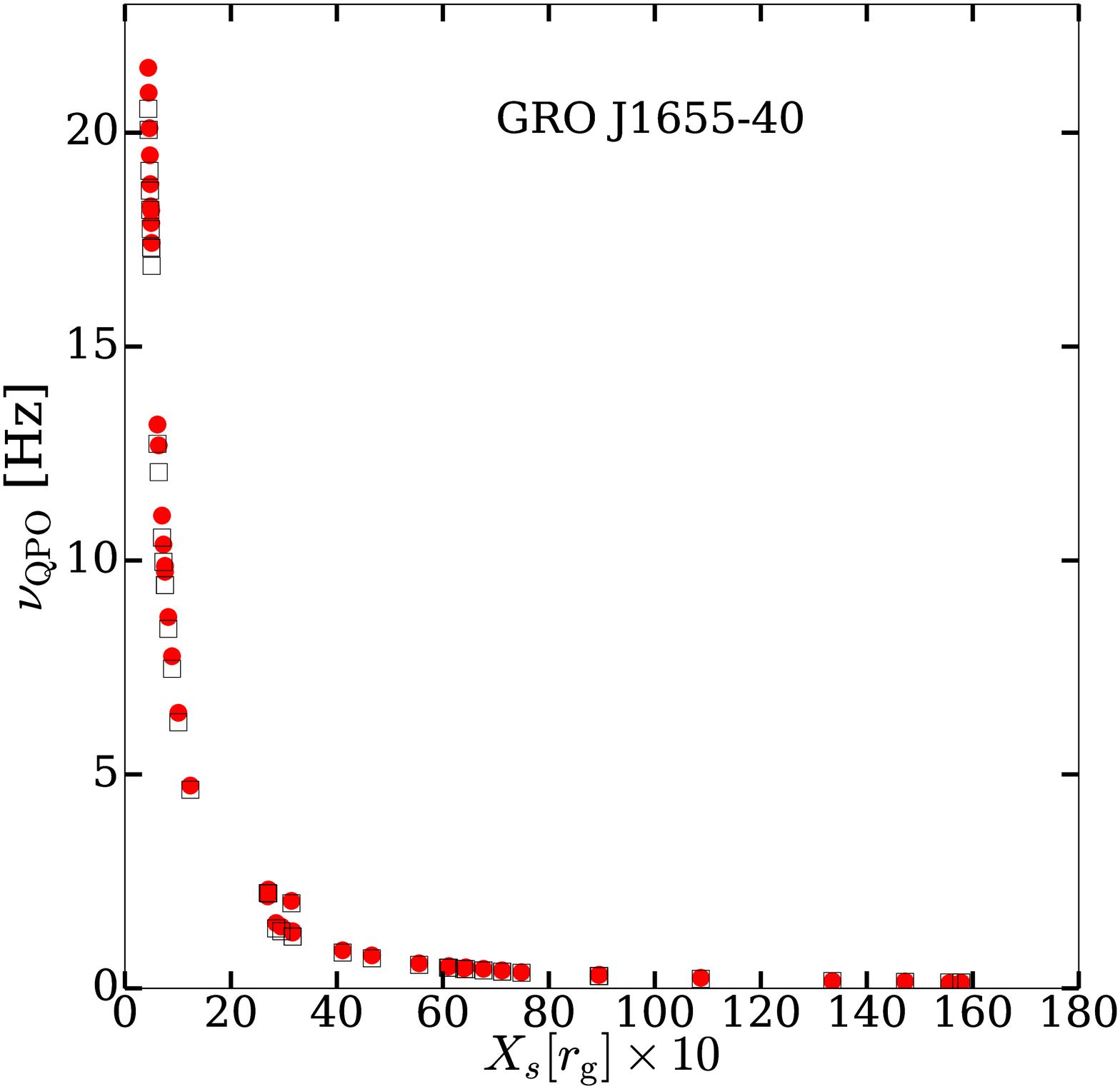}
\hspace{-0.6cm}
\includegraphics[width=0.37\linewidth]{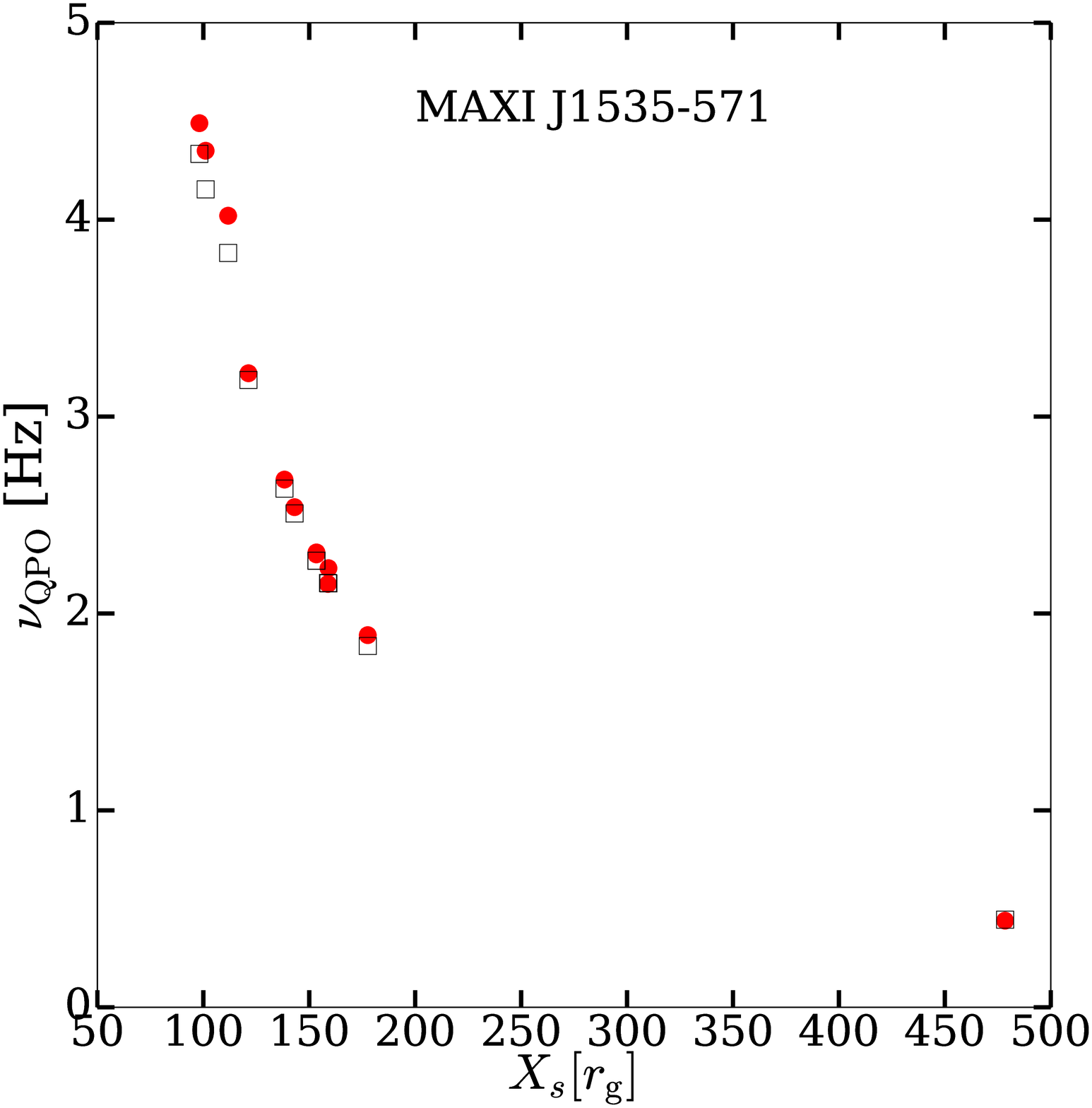}\\
\hspace{-1.2cm}
\includegraphics[width=0.37\linewidth]{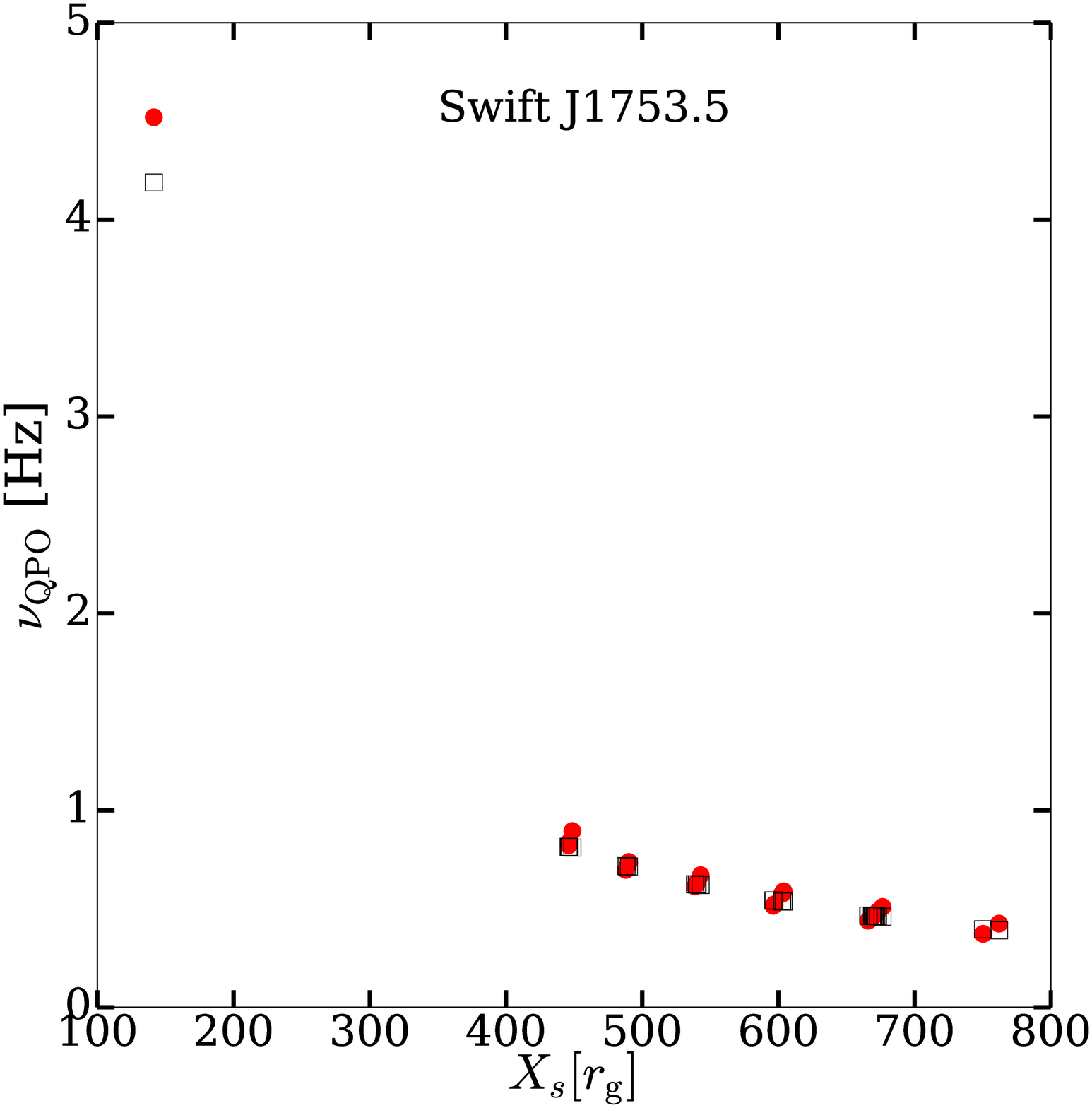}
\hspace{-0.6cm}
\includegraphics[width=0.37\linewidth]{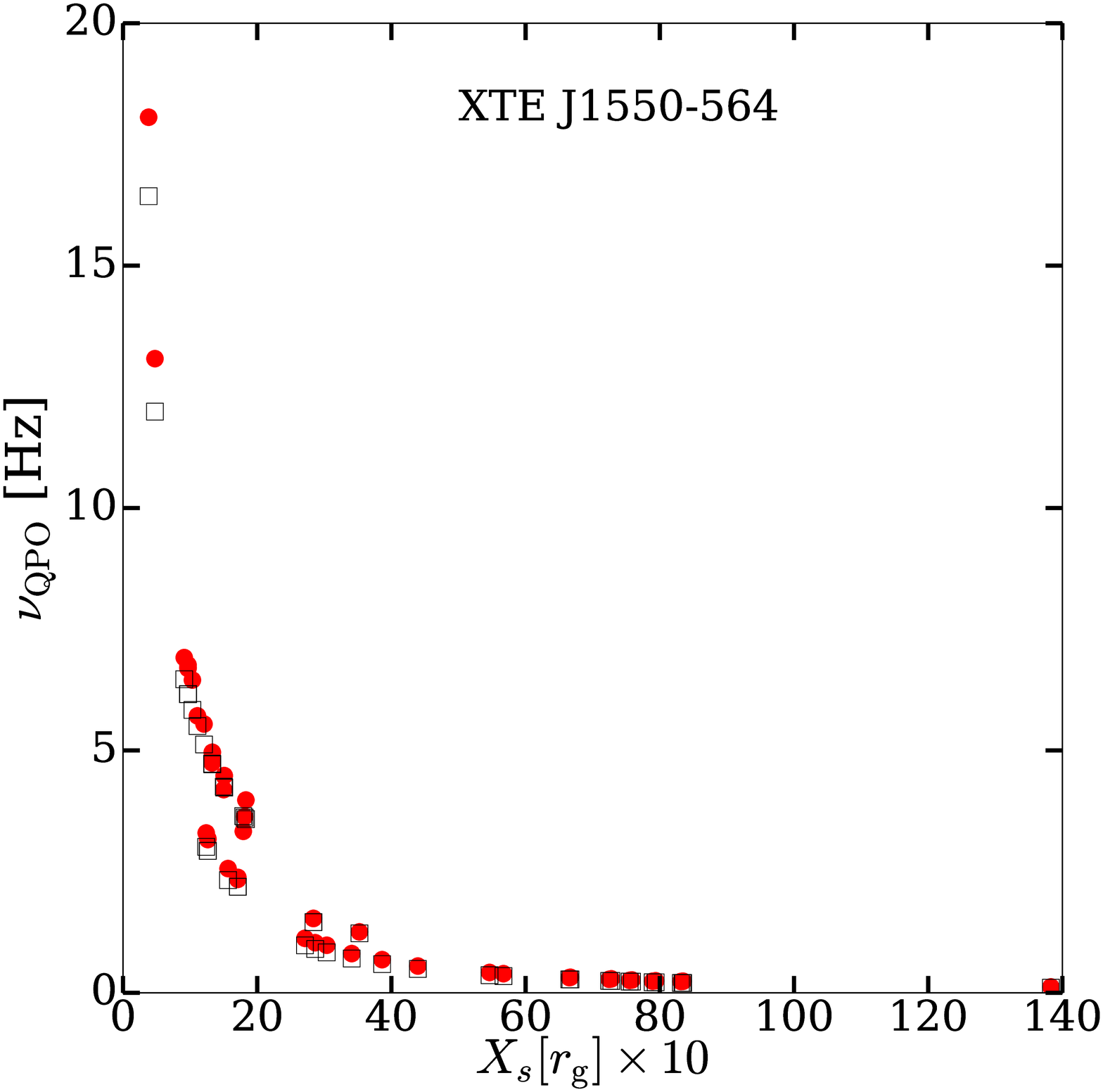}
\hspace{-0.6cm}
\includegraphics[width=0.37\linewidth]{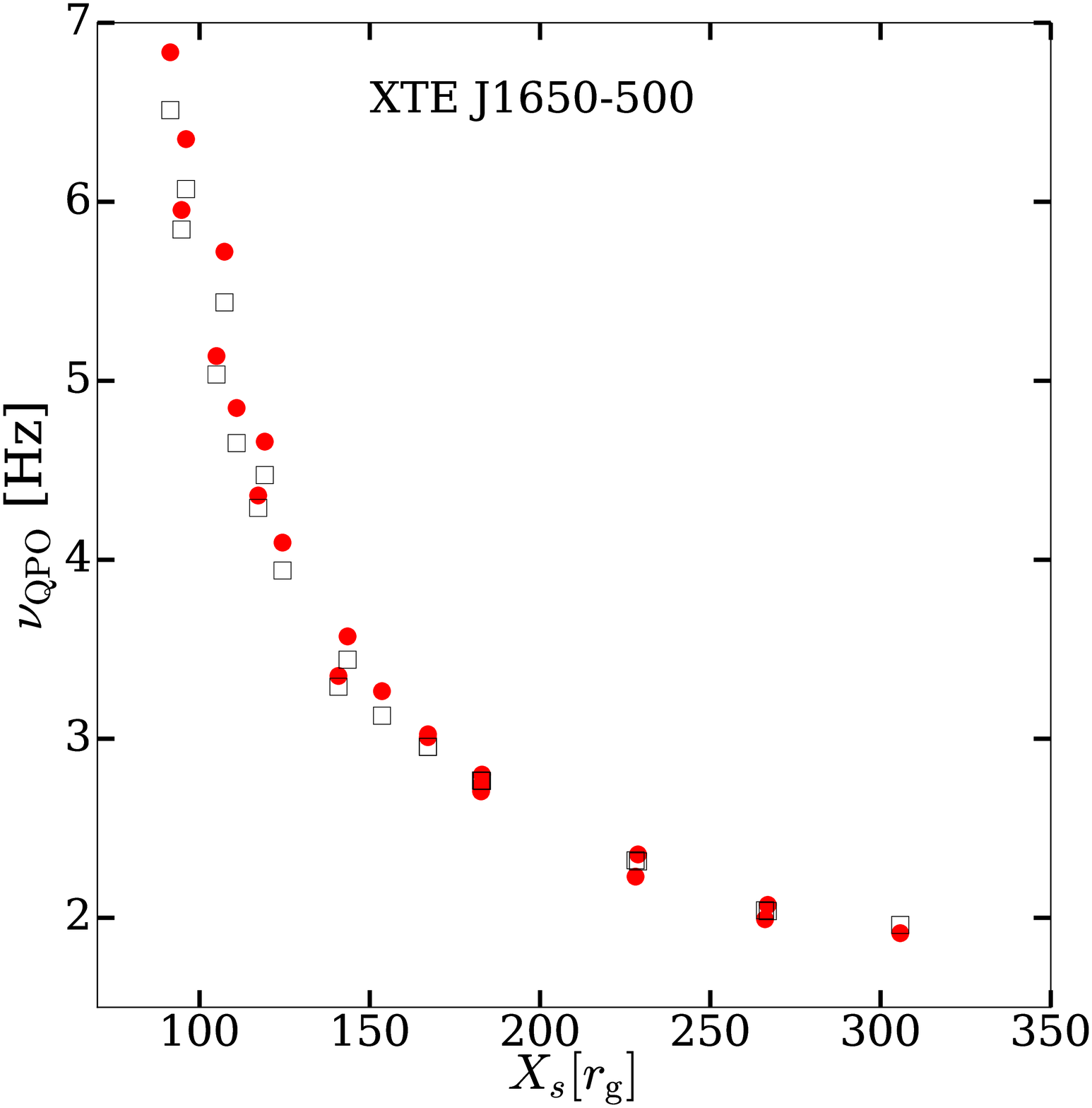}\\
\hspace{-1.2cm}
\includegraphics[width=0.37\linewidth]{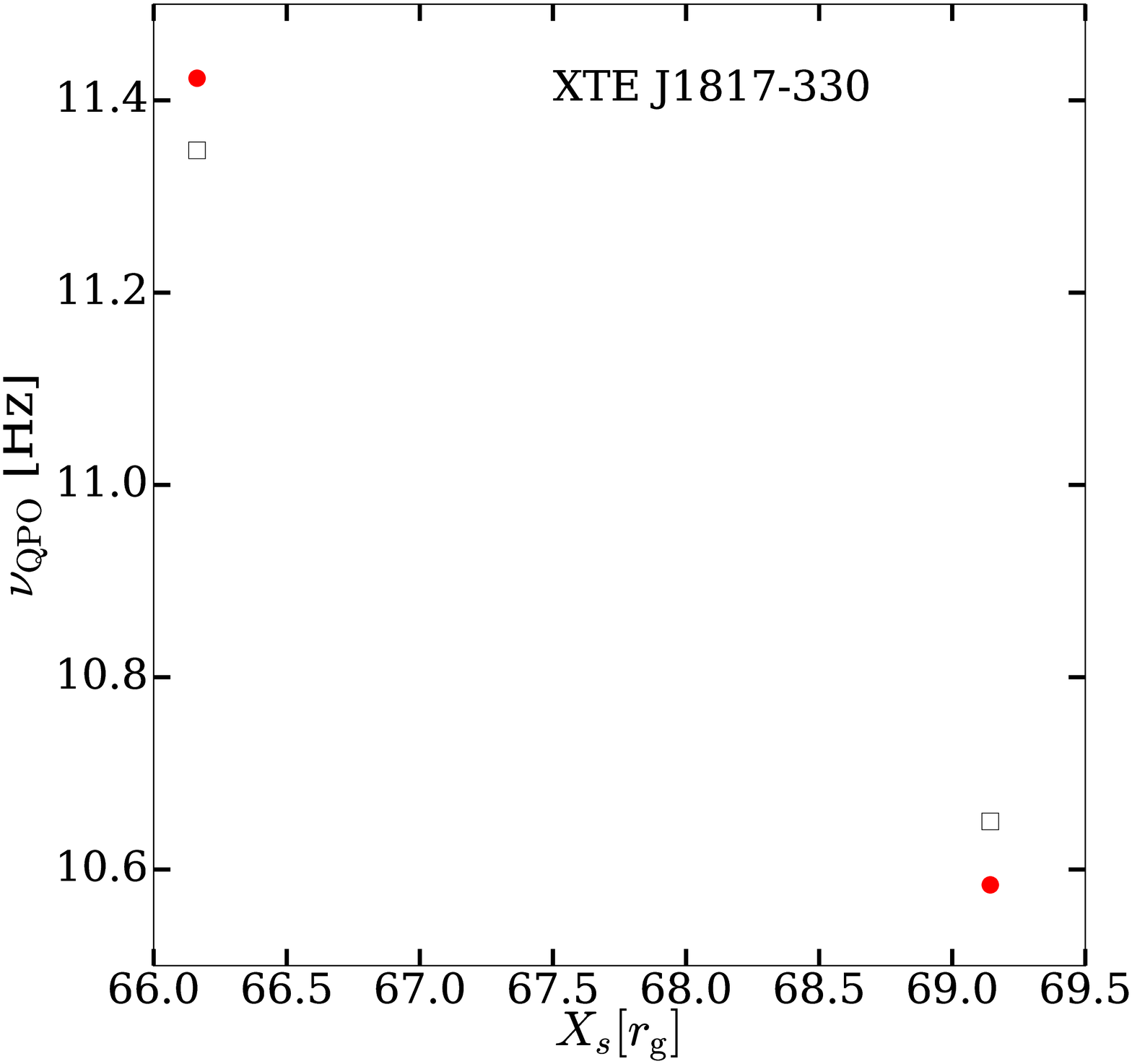}
\hspace{-0.6cm}
\includegraphics[width=0.37\linewidth]{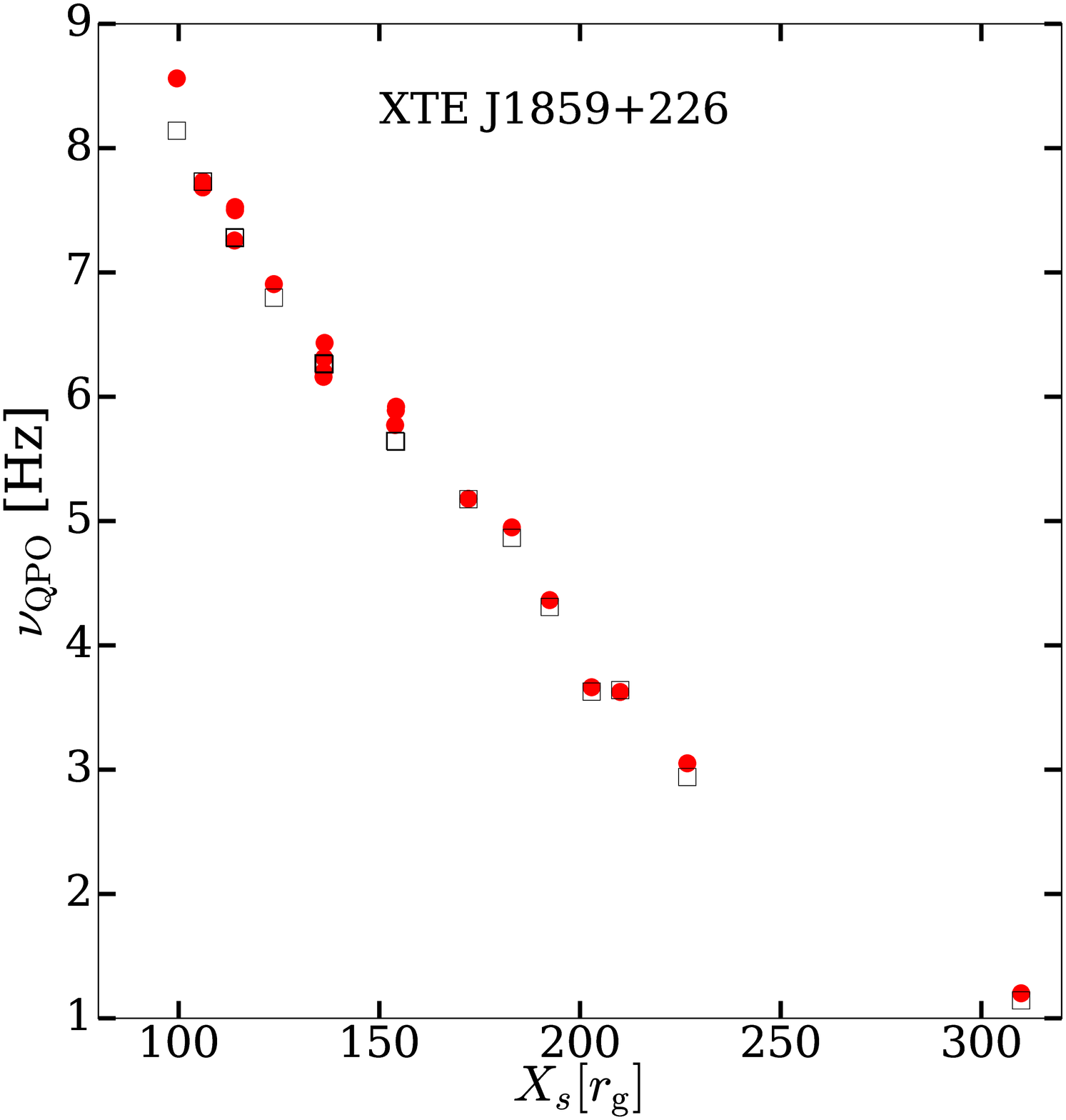}
}
\caption{QPO frequency, $\nu_{\rm QPO}$, evolution with shock location, $X_{s}$, for all BHXRBs with $M_{\rm BH}$ $< 10 M_\odot$. The solid red dots show the observed frequencies from Motta et al. (2015) and empty squares show the theoretically estimated frequencies. \label{fig:qpoVsXsMless10} Different empty squares are associated with the range of values of both $\varepsilon$ and $l$ as shown in table~\ref{table:results}}
\end{figure*}

The variation of $\nu_{\rm QPO}$ with $X_{s}$ for all BHXRBs with $M_{\rm BH}$ $< 10 M_\odot$ is shown in Fig.~\ref{fig:qpoVsXsMless10}. Using the range of values for parameters, $\varepsilon$ , $l$, $M_{\rm BH}$, and $a$, mentioned in table~\ref{table:results} for each BHXRB, we could successfully explain the observational data points (solid red dots) using our theoretical solutions (empty black squares). 
It can be concluded that to have higher values of $\nu_{\rm QPO}$, the shock needs to form closer to the BH and the corona region should be more compact (see Appendix \ref{App:hfqpos}). Similarly, Fig.~\ref{fig:qpoVsXsMGe10} shows results for a few BHXRBs with $M_{\rm BH}$ $\ge 10 M_\odot$. The theoretical estimate following the methodology in the previous section and using parameters mentioned in table~\ref{table:results}, the origin of observed values of $\nu_{\rm QPO}$. For all possible values of $M_{\rm BH}$, corona size seems to be of similar size to produce same $\nu_{\rm QPO}$. For the lowest value of $\nu_{\rm QPO}$ around 1 Hz, the dissipative standing shocks form at several hundreds of $r_{g}$ while for the highest value like around a few tens Hz the shock should be located at some tens of $r_{g}$. The general trend of the evolution of QPOs with shock location can be related to the spectral states change. As the state changes from a hard to soft through intermediate, the shock is located far away from the BH and moves inward, and the QPOs frequency gradually increases. Some sporadic QPOs are observed in intermediate states. These features have been observed for other BHXRBs that can be explained using hardness intensity diagram (Belloni 2010; Katoch et al. 2021; Alabarta et al. 2021; Shui et al. 2021 and references therein) or from mass accretion rates ratio diagram (Mondal et al. 2014).\\

\begin{figure*}
\centering{
\includegraphics[width=0.5\linewidth]{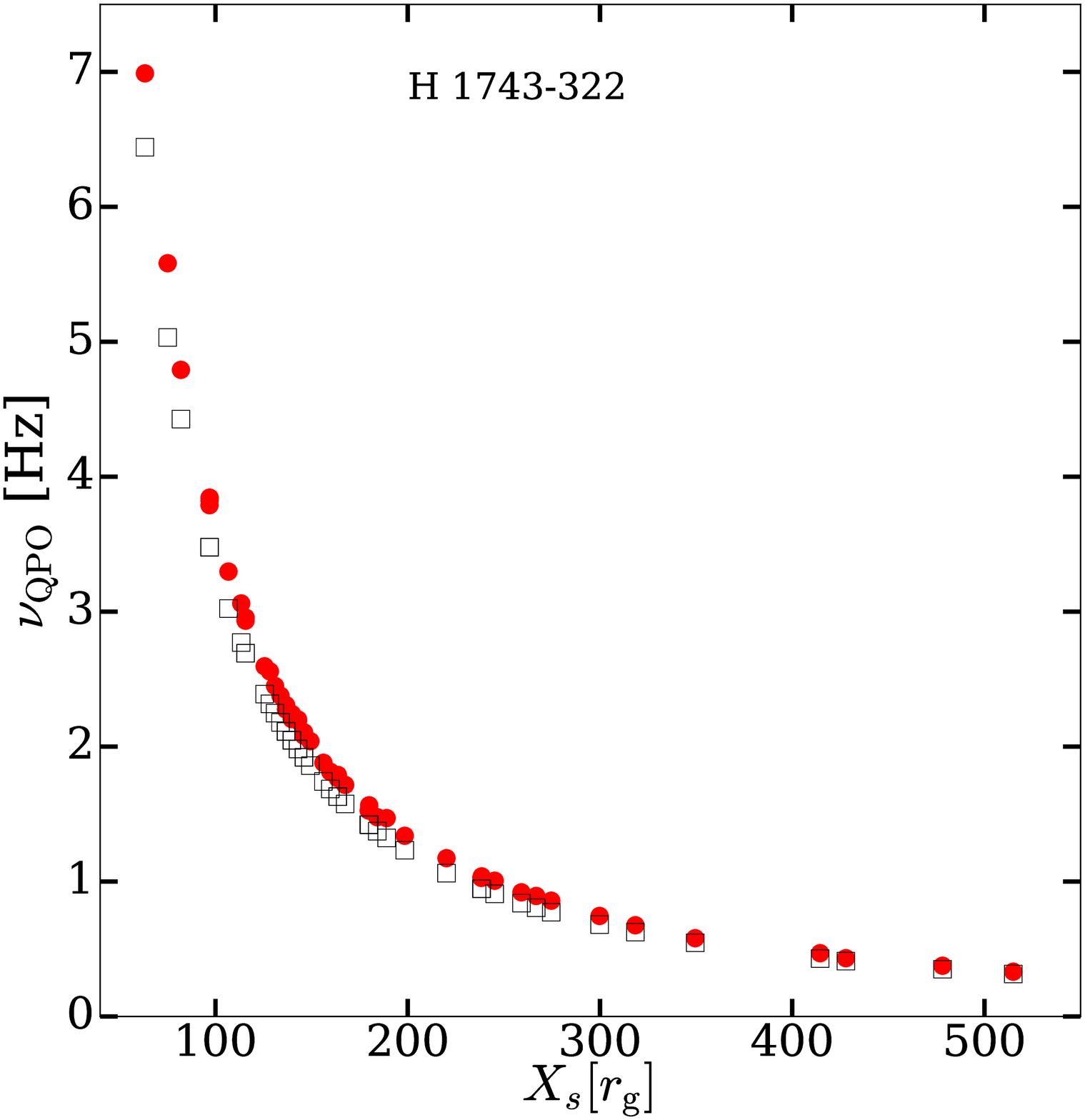}
\hspace{-0.5cm}
\includegraphics[width=0.5\linewidth]{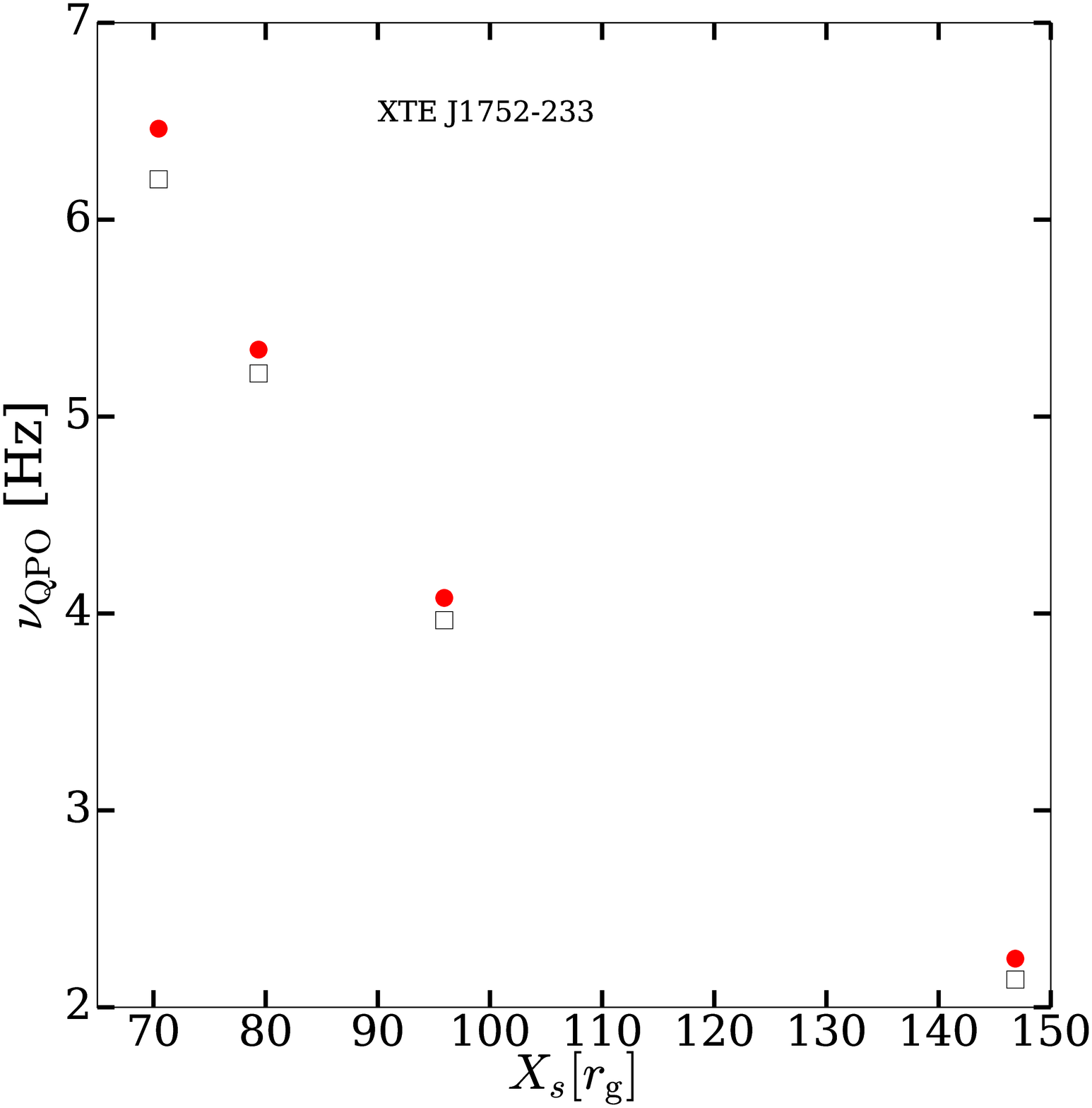}}
\caption{Same as \autoref{fig:qpoVsXsMless10}, but for candidates with $M_{\rm BH}$ $\ge 10 M_\odot$. \label{fig:qpoVsXsMGe10}}
\end{figure*}

\begin{figure}
\centering{
\includegraphics[width=1.1\linewidth]{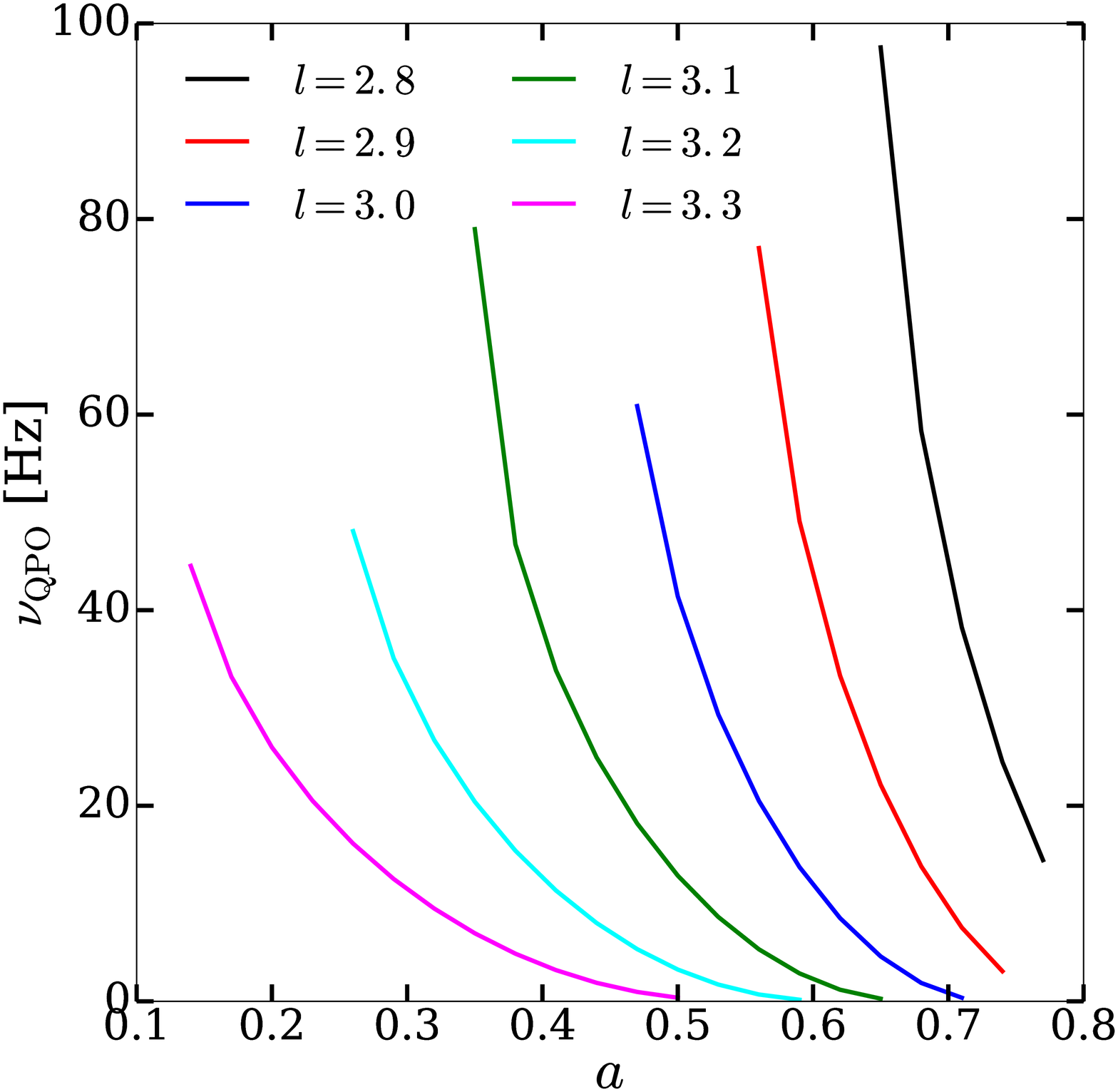}}
\caption{Dependence of QPO frequency, $\nu_{\rm QPO}$, on the BH spin, $a$. Different curves correspond to a fixed value of $l$. The other flow parameters are set fixed at $\varepsilon=1.0001$, $M_{\rm BH}=10 M_\odot$, and $\Delta \varepsilon=4\times10^{-5}$. \label{fig:SpinVsQPO}}
\end{figure}

\begin{figure}
\centering{
\includegraphics[width=1.1\linewidth]{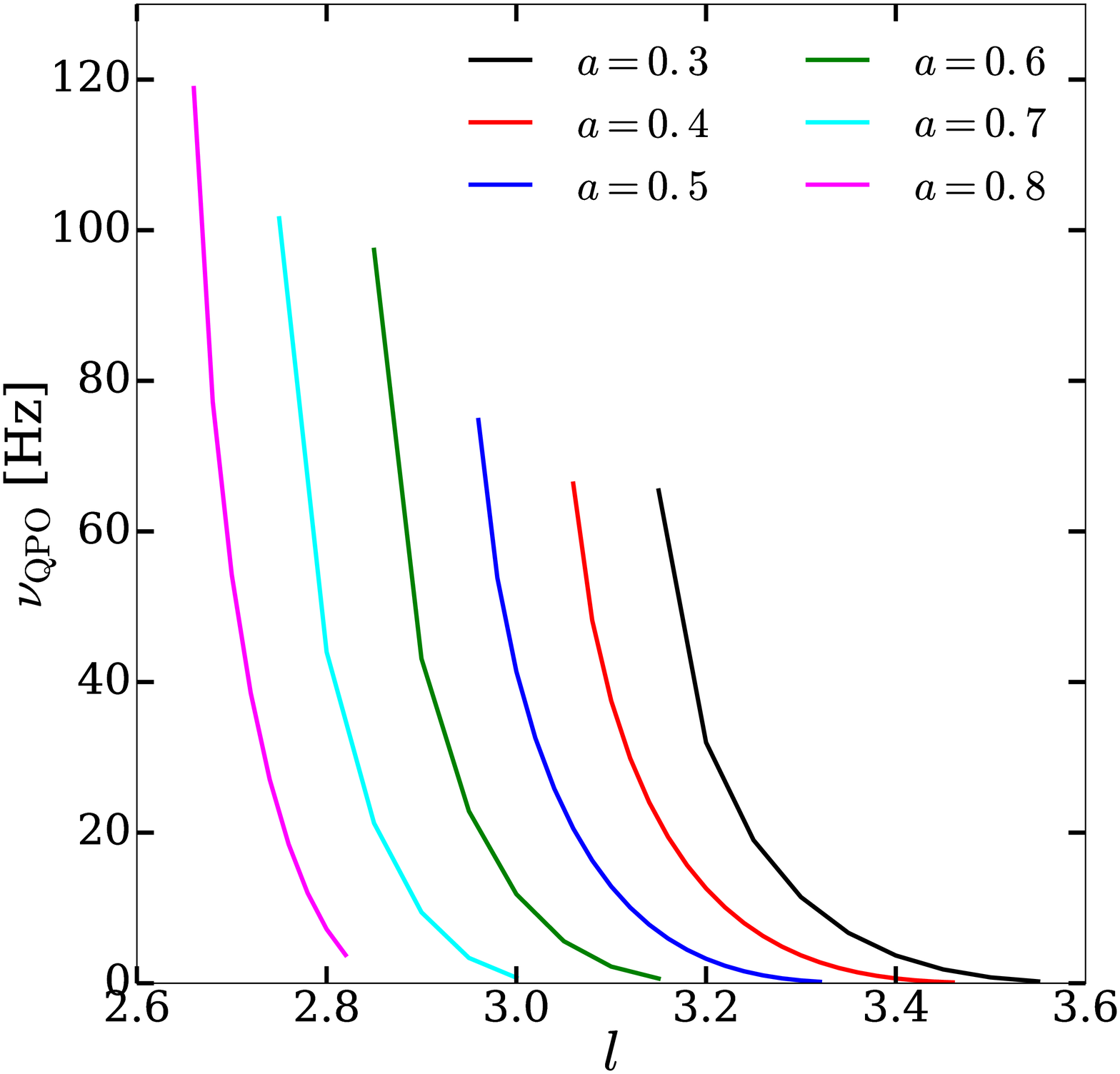}}
\caption{Variation of QPO frequency, $\nu_{\rm QPO}$, with specific angular momentum of the flow, $l$. Different curves correspond to a fixed value of $a$. The other flow parameters are set fixed at $\varepsilon=1.0001$, $M_{\rm BH}=10 M_\odot$, and $\Delta \varepsilon=4\times10^{-5}$. \label{fig:AngVsQPO}}
\end{figure}

Fig.~\ref{fig:SpinVsQPO} shows the theoretical estimate of $\nu_{\rm QPO}$ versus BH spin, $a$, for different values of $l$. Some parameters like $\varepsilon$, $\Delta\varepsilon$ and $M_{\rm BH}$ have been assigned particular values like $1.0001$, $4\times10^{-5}$ and $10 M_{\odot}$ respectively. The peak values of $\nu_{\rm QPO}$ are found to be the largest for highest $a$ and lowest $l$. For particular $a$, $\nu_{\rm QPO}$ increases with the decrease in $l$ while it decreases with an increase in $a$ for particular $l$. The dependence of $\nu_{\rm QPO}$ on $l$ for particular values of $a$ are shown in Fig.~\ref{fig:AngVsQPO}. The behaviour of $\nu_{\rm QPO}$ in Fig.~\ref{fig:AngVsQPO} follows the same trend as in Fig.~\ref{fig:SpinVsQPO}. The reason behind the above dependencies is as follows: higher spin parameter pushes the shock location away from the BH and hence decreases the infall velocity of the flow ($v_-$), so based on the equation of $t_{\rm infall}$, it increases thus decreases  $\nu_{\rm QPO}$. Similar behaviour is observed in $l$ v/s $\nu_{\rm QPO}$ variation.

\section{Summary and conclusions}
In this paper we address the physical origin of type-C LFQPOs in almost all BHXRBs observed from RXTE data. The oscillation of standing shock caused by the resonance condition due to matching of cooling and infall timescales can successfully explain the scenario. This scenario can also explain the spectral states. It is in this two-fold ability of its explanation that we see its power. At the onset of the outburst, the Keplerian disc just starts forming far from the BH, when only the hot sub-Keplerian flow dominates and the spectrum becomes harder, as the disc proceeds inward, more soft photons intercept hot electrons in the corona, therefore cooling increases and corona shrinks. When the cooling is significant enough to destruct the corona completely, the disc becomes fully Keplerian, which is the SS or can be high SS. In between these HS and SS, when the cooling timescale matches with the compressional heating timescale $-$ the resonance condition is fulfilled, the QPO appears (Molteni et al. 1996, Garain et al. 2014). Applying the above theory to large samples, our main results are as follows:\\

(1) There is a power-law dependence of cooling factor in the post-shock region on the QPO frequency which is possibly generic in nature, requires further study.

(2) The thermal Comptonisation cooling leads to compact corona forming close to the BH and a smaller standing shock location causes the higher value of QPO frequency. So the standing shock location is correlated with the QPO frequency. 

(3) The QPOs due to oscillation of the standing shocks are likely to be weakly dependent on BH masses and spins as the shock forms far from the BH.

(4) Only two free parameters of the accretion flow like specific angular momentum and specific energy are sufficient to explain the QPO frequencies in all given BHXRBs.

In the present work, for mathematical simplicity pseudo-Kerr potential has been invoked to describe the BH space-time however a full general-relativistic study has been left for future work and will be reported elsewhere. The estimation of cooling energy from the power-law spectral component for all sources in Table~\ref{table:results} is beyond the scope of this paper. We aim to estimate the above cooling and verify our present claim of $f_c$ in follow-up work.

\section*{Acknowledgments}
CBS  is  supported  by  the  National  Natural  Science  Foundation  of China under grant No. 12073021. SM acknowledges Ramanujan Fellowship grant (\#RJF/2020/000113) for this research.

\section*{Data Availability}
Data information may not be applicable for this article. No new data has been analysed, as it is mostly a theory-based article. The numerically created data that support the findings of this study are available from the corresponding authors, upon reasonable request.

\appendix
\section{Representative cases for HFQPOs} \label{App:hfqpos}
We have estimated HFQPOs for two BHXRBs with mass $<10 M_\odot$ (GRO~J1655-40; as J1655 in table ~\ref {table:HFQPOsresults}) and $>10 M_\odot$ (H\,1743-322) as representative cases. We explore the parameter space of $\varepsilon$ and $l$ and the theoretical solutions give the permissible values of $X_{s}$ and $R$ corresponding to only two sets of $\varepsilon$ and $l$ values as shown in table~\ref {table:HFQPOsresults} and can explain the origin of HFQPO frequencies compatible with the observations. For GRO~J1655, 300~Hz HFQPO (Remillard et al. 1999; Strohmayer 2001) can be originated when the shock is much closer to the BH at $10.3$ or $11.4 r_g$ and for H\,1743, 166~Hz HFQPO (Homan et al. 2005; Remillard et al. 2006) can be originated when the shock is at $11$ or $13 r_g$ depending on the values of $\varepsilon$ and $l$. We should mention that for HFQPOs, the degeneracy is less due to the shrinking of the allowed shock-forming region. We also mention that though $X_s$ and R are not very different, the very different QPO frequencies for the two sources are due to the mass of the BH, as it also scales the QPO frequency inversely.

\begin{table}
\centering
\caption{OBSERVED AND THEORETICAL PARAMETERS USED FOR ESTIMATING THEORETICAL HFQPOs FOR TWO REPRESENTATIVE SOURCES.  \label{table:HFQPOsresults}}
\begin{tabular}{lcccccr}
\hline
Source     &$M_{\rm BH}$ &$a$ &$\varepsilon$ &$l$&$X_s$  &R\\
\hline
J1655      &6.0    &0.5    &1.0089, 1.0118 &2.69, 2.75&10.3, 11.4  &2.4, 1.9\\
H\,1743    &11.2   &0.4    &1.01, 1.0144 &2.68, 2.79&11, 13  &2.1, 1.4\\
\hline
\end{tabular}
\\
\end{table}

\end{document}